\documentclass[useAMS,usenatbib]{mn2e}
\usepackage{rotating}

\title[Asteroid rotation excitation by subcatastrophic impacts]{Asteroid rotation excitation by subcatastrophic impacts}
\author[Tom\'{a}\v{s} Henych, Petr Pravec]{Tom\'{a}\v{s} Henych$^{{1},{2}}$\thanks{E-mail:ftom@physics.muni.cz (TH)}, Petr Pravec$^{2}$
\\
$^{1}$Dept. of Theoretical Physics and Astrophysics, Faculty of Science, Masaryk University, Kotl\'{a}\v{r}sk\'{a} 2, CZ-61137 Brno, Czech Republic\\
$^{2}$Astronomical Institute, Academy of Sciences of the Czech Republic, Fri\v{c}ova 298, CZ-25165 Ond\v{r}ejov, Czech Republic}

\begin{document}
\date{Accepted 2013 April 2 \hspace*{3em} Received 2013 March 27; in original form 2013 February 8}

\pagerange{\pageref{firstpage}--\pageref{lastpage}} \pubyear{2013}

\maketitle

\label{firstpage}

\begin{abstract}
Photometric observations of asteroids show that some of them are in non-principal axis rotation state (free precession), called tumbling. Collisions between asteroids have been proposed as a possible asteroid rotation excitation mechanism. We simulated subcatastrophic collisions between asteroids of various physical and material parameters to find out whether they could be responsible for the excited rotation. For every simulated target body after the collision, we computed its rotational lightcurve and we found that tumbling was photometrically detectable for the rotational axis misalignment angle $\beta$ greater than about $15^\circ$. We found that subcatastrophic collisions are a plausible cause of non-principal axis rotation for small slowly rotating asteroids. The determining parameter is the ratio of the projectile orbital angular momentum to the target rotational angular momentum and we derived an approximate relation between this ratio and the angle $\beta$. We also compared the limiting energy for the onset of tumbling with the shattering energy. Slowly rotating asteroids of diameter 100~m and larger can be rotationally excited by collisions with energies below the shattering limit.

\end{abstract}


\section{Introduction}\label{intro}

Most asteroids appear to be in a basic state of rotation, i.e., they rotate around the shortest principal axis (the axis with the maximum moment of inertia). We know, however, that asteroids with small sizes and slow rotations are in non-principal axis rotation states, for which~\cite{harris1994} coined the term \emph{tumbler}. They show lightcurves with two frequencies (and their linear combinations), reflecting their rotational state of free precession. The first confirmed tumbler was 4179~Toutatis (\citealt{spencer1995}, \citealt{ostro1995}), a small Apollo asteroid that was thoroughly observed in December 1992 around its close approach to Earth.

Since the non-principal axis rotation state has a higher energy than the basic one, we are interested in what processes excite asteroid rotations. Proposed explanations were a solar radiation pressure toque (\citealt{paddack1969}), a torque given by absorption of sun radiation and its re-emission as thermal radiation by irregular asteroid surface with a thermal inertia (the Yarkovsky--O'Keefe--Radzievskii--Paddack, or YORP, effect; \citealt{rubincam2000}, \citealt{vokrouhlicky2007}), ejecta escaping from the target asteroid after the collision (\citealt{scheeres2000}), gravitational torques during close planetary flybys (only possible for planet-crossing asteroids, \citealt{richardson1998}, \citealt{black1999}), and a torque caused by mass ejection from a comet (\citealt{peale1989}). More discussion on this can be found in~\cite{paolicchi2002}.

~\cite{pravec2005} presented a review of known tumblers, their dynamical characteristics and also speculated on the origin of their spins. They discussed the YORP effect as a process that could be responsible for slowing down the rotation of asteroid and possibly even for their rotational excitation. The other possibility mentioned was a subcatastrophic collision which could be much more effective for an asteroid previously slowed down by YORP. 

Collisions between asteroids are a fundamental process, affecting asteroid spins and size distribution (\citealt{harris1979a}, \citealt{harris1979b}; \citealt{cellino1991}), forming asteroid families (\citealt{hirayama1918}, \citealt{hirayama1927}) and some binaries (\citealt{noll2006}), and also triggering activity of main belt asteroids (\citealt{jewitt2011}, \citealt{bodewits2011}; \citealt{stevenson2012}). It is interesting to ask, if collisions can be plausible explanation for the excitation of asteroid rotation. 

The most common surface features of the minor bodies with rocky surface are impact craters. Some craters are of substantial size. Actually, the largest craters observed on asteroids visited by space probes have $D/R_{\rm m}\sim 1$ and some even $D/R_{\rm m}\sim 1.5$, where $D$ is the crater diameter and $R_{\rm m}$ is the impacted body mean radius~--~the radius of a sphere of the same volume (\citealt{thomas1999b}). For instance, the tumbling asteroid 253~Mathilde with the mean radius of $26.4\,\rm{km}$ (\citealt{tedesco1992}) and very slow rotation (main period $418\,\rm{h}$, \citealt{mottola1995}), is covered by several huge impact craters. The largest one has a diameter of $33\,\rm{km}$ (\citealt{veverka1997}). 

The process of collisional excitation of asteroid rotation has been examined in some earlier papers (\citealt{gauchez2006}), but they did not specifically focus on the onset of tumbling. We use a similar parametrization of the problem, but we only assume explosive cratering events and concentrate on non-principal axis rotation.

In section~\ref{model}, we describe our numerical model that we used to study the effects of collisions between asteroids on their rotations, section~\ref{rot_axis_mis} shows the dependence of a rotational axis misalignment on physical and dynamical parameters of the colliding bodies, in section~\ref{am_ratio} we describe how the misalignment depends on the ratio of projectile's orbital AM to the target's rotational AM and in section~\ref{disrupt_crit} we plot specific impact energy of the collisions as a function of the mean target radius.

\section{Subcatastrophic Impact Model}\label{model}
We investigate the effect of subcatastrophic impact on the rotation of an asteroid in a following way: there are two bodies described by a set of physical, dynamical and shape parameters (size, density, velocity etc.) that experience hypervelocity collision. The projectile is completely destroyed and an impact crater is created on the target. We then recalculate the dynamical characteristics of the target, calculate the rotational lightcurve and check if the tumbling of the target can be detected by photometric observations. We describe our model in more detail in following.

The target is a homogeneous\footnote{In most of our simulations, material constants used involve some microporosity. In some simulations we modelled macro\-po\-ro\-si\-ty as well.} triaxial ellipsoid in relaxed rotation state, i.e., rotating around the shortest axis. It is described by following parameters: the mean radius of the body, the semi-axes ratios $a/c$ and $b/c$ ($a\geq b\geq c$), the body's bulk density $\rho$, the material strength\footnote{We assume effective strength for the excavation, because the shock wave fractures the body during the early phase of the impact (it is not tensile, shear or other material strength that can be measured in laboratory); for details see \cite{richardsonj2009}, \cite{nolan1996} and \cite{holsapple2007}.}, initial rotation period and whether the body is porous or not (see chapter~\ref{scaling} for details on this). The projectile is a homogeneous sphere with diameter $d$ and the mean density $\delta$. We do not take into account the rotation of the projectile since its angular momentum is negligible.

We observe the impact in an inertial frame connected with the centre of mass of the two bodies. The frame coordinate axes are identical to the principal axes of the target at the moment of impact ($x$ in the direction of the target's longest axis, $z$ in the direction of the shortest one).

The projectile is moving towards the target at a relative velocity $\bmath{v}$ and hits it at the impact point $I$ with spherical coordinates $\phi$ (longitude) and $\theta$ (latitude) on the surface of the target body. The prime meridian is defined as $y=0$ and $x\geq0$; together with the longest target axis it makes the common plane.

The impact speed is much greater than the target's escape velocity so we can neglect mutual gravity of the bodies and any curvature of the projectile's trajectory (\citealt{love1996}).

The projectile is completely destroyed in the impact and a crater is formed on the surface of the target at the impact point. Its diameter and depth are calculated by using scaling laws based on point source approximation of the impact (\citealt{holsapple1993}; \citealt{holsapple_housen2007}).

The geometrical representation of the crater is a paraboloid of revolution; this corresponds to the observations of simple impact craters on the Moon and other bodies in the Solar System (\citealt{chappelow2002}). The axis of the paraboloid is perpendicular to the local surface. We do not take into account border rim of the crater (which is usually present in real craters); we only model simple craters in our simulations. Complex impact structures can only be observed on the largest asteroids in the main belt and on large icy bodies in the outer Solar System where the gravity plays a substantial role in crater formation (\citealt{leliwa2008}).

The diameter of the largest crater that a body of a given size can bear is calculated according to the relation in \citet{burchell2010}. They studied the maximum crater size on small bodies (asteroids, icy satellites, comets), which were imaged by space probes, and showed, that it follows the relation
\begin{eqnarray}
D=-(0.17\pm0.10)+(1.01\pm0.08)R_{\rm m}\,\nonumber\\
0.7\,{\rm km}<R_{\rm m}<120\,{\rm km}\,,
\end{eqnarray}
$D$ being crater diameter and $R_{\rm m}$ the target mean radius both in kilometres. 

The relative diameters of the impact craters $F=D/R_{\rm m}$, formed on the target in our simulations, were usually kept less than 1.26, which is the value for the largest crater on 253~Mathilde (\citealt{leliwa2008}). In some simulations we allowed for higher values of up to $F=1.6$, since there is indirect evidence of large crater on 90~Antiope (\citealt{descamps2009}). However, conclusions based on this higher limit should be made with caution.

For every collision we calculated also its specific impact energy (kinetic energy of the collision per unit mass of the target) and compared it to the dispersal criterion as described by \cite{stewart2012}. We plot our simulation results in a graph of specific impact energy vs. mean target radius and discuss the results in section~\ref{disrupt_crit}.

The ejecta are assumed to leave the impact point with a cone-shaped velocity field symmetric around the line perpendicular to the surface at the impact point. We take this effect into account when calculating the angular momentum transfer efficiency.

\subsection{Inertia tensor calculation}\label{num_meth}

The inertia tensor of the target before the impact is calculated analytically for the body being homogeneous triaxial ellipsoid. The inertia tensor after the impact is calculated numerically since the body is irregular\=it is the triaxial ellipsoid with the crater on its surface. 

The body surface is divided into triangular facets and together with the origin they form the tetrahedron. Its inertia tensor is calculated analytically (see, e.g., \citealt{tonon2004}) and using the Steiner's parallel axis theorem it is translated into the original centre of mass coordinate system (reference frame, origin $O$). The inertia tensors of all such tetrahedra are then summed up and the inertia tensor of whole body is obtained. Then it is translated to the new reference frame with origin $O^\star$, since the centre of mass has shifted due to the impact.

The situation with facets covering the crater is only a bit more complicated. First the crater border is calculated. Then all the facets' vertices of the ellipsoid surface inside this circular border are replaced with points of the paraboloid describing the crater floor. New facets' vertices are made of these points and the same procedure of tetrahedra inertia tensor calculation is used then.

\subsection{Principal axis deviation calculation}\label{deviation}

We try to find out whether a subcatastrophic impact can create an observable excitation of rotation of the impacted body. The target's rotation was relaxed before the impact. It means it was rotating about the shortest axis, i.e., the axis with the largest moment of inertia for the homogeneous ellipsoid. We denote this moment of inertia $I_3$ and it holds that $I_1\leq I_2<I_3$. 

An indication of the excited rotation will be a nonzero angle between the rotational angular momentum vector of the body after the impact $\bmath{L^{\!\star}}$ and its shortest principal axis with the largest moment of inertia ${I_3}^{\!\!\star}$. The procedure of its calculation is as follows.

First the inertia tensor of the target after the impact is calculated. Since the body possesses the impact crater, it lacks symmetry and the inertia tensor is calculated numerically.

Because the centre of mass of the target $O^\star$ is shifted with respect to the original centre of mass $O$, the resulting inertia tensor has to be shifted accordingly (the method calculates it with respect to $O$, hence it is not a central inertia tensor for the new body). We want to have the inertia tensor in a body frame (which is not inertial) with an origin in its centre of mass, because then the components of this inertia tensor do not change with time as the body rotates. The transformation reads
\begin{equation}
  \mathbfss{I}_{ik}^{\star} = \mathbfss{I}_{ik} + M(a^2\delta_{ik}-a_ia_k)\,,
\end{equation}
where $M$ is the total mass of the body, $\bmath{a}$ is the centre of mass shift vector (between $O$ and $O^\star$), $a$ is its magnitude, $a_i$ and $a_k$ are its cartesian components and $\delta_{ik}$ is Kronecker delta (see, e.g., \citealt{kvasnica}). 

Now we have to find the principal axes of the target after the impact. We can calculate them by finding the diagonal form of its inertia tensor, because they are determined by the eigenvectors of this tensor. The diagonalization is given by
\begin{equation}
  \mathbfss{I}_{\rm D} = Q^{\rm T}\mathbfss{I}^\star Q\,,
\end{equation}
where $\mathbfss{I}_{\rm D}$ is diagonal inertia tensor, $Q$ and $Q^{\rm T}$ are orthogonal matrix and transposed orthogonal matrix, respectively. Its rows (and columns, respectively) are the eigenvectors of $\mathbfss{I}^\star$. 

In the next step we calculate the angular momentum vector just after the impact. We use the momentum conservation law
\begin{equation}
  m_{\rm t}\bmath{v_{\rm t}} + m_{\rm p}\bmath{v_{\rm imp}} = m^\star\bmath{v^\star} + \bmath{P_{\rm ejecta}}\,,
  \label{mom_conservation}
\end{equation}
and angular momentum conservation law
\begin{equation}
  \bmath{L_{\rm t}} + \bmath{L_{\rm p}} + \bmath{L_{\rm orb,t}} + \bmath{L_{\rm orb,p}} = \bmath{L^{\!\star}} + \bmath{L_{\rm orb}^\star} + \bmath{L_{\rm ejecta}}\,,
  \label{am_conservation}
\end{equation}
where $m_{\rm p}$ is the mass of the projectile, $\bmath{v_{\rm imp}}$ is the projectile velocity vector, which for very low mass ratio of the projectile to target approaches the impact velocity, $m_{\rm t}$ and $m^\star$ are the masses, $\bmath{v_{\rm t}}$ and $\bmath{v^\star}$ are the velocities of the target before and after the collision, respectively, and $\bmath{P_{\rm ejecta}}$ is the momentum of the ejecta. $\bmath{L_{\rm t}}$ and $\bmath{L_{\rm p}}$ are the rotational angular momenta of the target and the projectile before the impact, respectively, $\bmath{L_{\rm orb,t}}$ and $\bmath{L_{\rm orb,p}}$ are the orbital angular momentum vectors of the target and the projectile, respectively, with respect to the system's centre of mass, $\bmath{L^{\!\star}}$ and $\bmath{L_{\rm orb}^\star}$ are the rotational and orbital angular momenta of the target after the impact, respectively, and $\bmath{L_{\rm ejecta}}$ is the angular momentum of the material ejected during the impact. Since the projectile has dissintegrated during the impact, all the quantities on the right sides of Eqn.~\ref{mom_conservation} and~\ref{am_conservation} describe the postimpact target body, and so we omit the subscript $t$ .

The orbital angular momenta are calculated as
\[
  \bmath{L_{\rm{orb,p}}} = m_{\rm p}(\bmath{r_{\rm{p}}}\times\bmath{v_{\rm{imp}}})
\]
\begin{equation}
  \bmath{L_{\rm{orb,t}}} = m_{\rm t}(\bmath{r_{\rm{t}}}\times\bmath{v_{\rm{t}}})\label{orb_momenta}
\end{equation}
\[
  \bmath{L_{\rm{orb}}^\star} = m^\star(\bmath{r^\star}\times\bmath{v^\star})\nonumber
\]
where $\bmath{r_{\rm{p}}}$, $\bmath{r_{\rm{t}}}$ and $\bmath{r^\star}$ are the radius vectors of the projectile and the target before and after the impact, respectively, with respect to the system's centre of mass. We calculate the target velocity before the impact as
\begin{equation}
  \bmath{v_{\rm{t}}} = -\frac{m_{\rm p}}{m_{\rm t}}\bmath{v_{\rm{imp}}}\,.
  \label{targ_velocity}
\end{equation}
Without an appropriate model of the ejecta, we need to at least roughly estimate its dynamical effect. We decided to use the efficiency of the linear (and angular) momentum transfer measured in laboratory impact experiments by \citet{yanagisawa1996} and \citet{yanagisawa2000} and we assumed that this relationship holds also for large scale impacts. Using the efficiencies $\eta$ and $\zeta$, the linear momentum transfer efficiency in normal and tangential direction with respect to the local surface, respectively, we calculated the $\bmath{v^\star}$ including the effect of ejecta. For the definition of these efficiencies and more detailed discussion see section~\ref{efficiency}.

Lastly we needed to describe $\bmath{L_{\rm ejecta}}$. For small cratering impact we can calculate the difference $\bmath{L_{\rm{orb,p}}} - \bmath{L_{\rm ejecta}}$ as
\begin{equation}
  \bmath{L_{\rm{orb,p}}} - \bmath{L_{\rm ejecta}} = \eta m_{\rm p}(\bmath{r_{\rm{p}}}\times\bmath{v_{\rm norm}}) + \zeta m_{\rm p}(\bmath{r_{\rm{p}}}\times\bmath{v_{\rm tang}})\,,
  \label{orb_momentum}
\end{equation}
where $\bmath{v_{\rm norm}}$ and $\bmath{v_{\rm tang}}$ are impact velocity components normal to and tangential to the local surface of the target, respectively. 

The rotational angular momenta of the bodies are simply given by
\begin{equation}
  \bmath{L_{\rm t}} = \mathbfss{I}_{\rm t}\bmath{\omega_{\rm t}}\qquad\hbox{and}\qquad\bmath{L_{\rm p}} = \mathbfss{I}_{\rm p}\bmath{\omega_{\rm p}}\,,
  \label{rot_momenta}
\end{equation}
where we introduced inertia tensors of the target $\mathbfss{I}_{\rm t}$ and of the projectile $\mathbfss{I}_{\rm p}$ and angular velocities $\bmath{\omega_{\rm t}}$ and $\bmath{\omega_{\rm p}}$ of the target and the projectile, respectively.

When we put the equations \ref{orb_momenta}, \ref{targ_velocity}, \ref{orb_momentum} and \ref{rot_momenta} to the equation of angular momentum conservation \ref{am_conservation}, we can calculate the angular momentum vector of the target after the impact as
\begin{equation}
  \bmath{L{\!^\star}} = \mathbfss{I}_{\rm t}\bmath{\omega_{\rm t}} + \mathbfss{I}_{\rm p}\bmath{\omega_{\rm p}} + (\bmath{L_{\rm{orb,p}}} - \bmath{L_{\rm ejecta}}) + \bmath{L_{\rm{orb,t}}} - \bmath{L_{\rm{orb}}^\star}\,.
\end{equation}
 
Finally we can calculate the angle between the rotational angular momentum vector of the target after the impact $\bmath{L^{\!\star}}$ and the axis with the largest moment of inertia $I_3^\star$ (the axis is represented by unit vector $\bmath{E_3}$, the eigenvector of the inertia tensor $\mathbfss{I}^\star$). We denote this angle as $\beta$,  
\begin{equation}
  \beta = \arccos\left(\frac{\bmath{L^{\!\star}}\bmath{E_3}}{|\bmath{L^{\!\star}}||\bmath{E_3}|}\right)\,.
\label{beta_main}
\end{equation}

The angle $\beta$ is not constant, but it varies with changes of the principal vector $\bmath{E_3}$ as the body precesses. We calculate the value of $\beta$ right after the collision and we use it as an approximate measure of the magnitude of tumbling displayed in lightcurve, for which the use of the instantaneous value of the misalignment angle is sufficient.

\subsection{Angular Momentum Transfer Efficiency}\label{efficiency}

During the collision the projectile angular momentum is transferred to the target. The target body, however, does not obtain all of the angular momenta carried by the projectile, as shown by the experiments of small scale impacts (see, e.g., \citet{yanagisawa1996}, \citet{yanagisawa2000} and references therin). A part of the angular momentum is carried away by high velocity ejecta and spall fragments and so we introduce the momentum transfer efficiency\footnote{The efficiency can be larger than one because backfire ejecta carries away some momentum in the direction opposite to the projectile course.} defined as
\begin{equation}
  \eta = \frac{m^\star v_{\rm norm}^\star}{m_{\rm p}v_{\rm norm}}\qquad{\rm and}\qquad\zeta = \frac{m^\star v_{\rm tang}^\star}{m_{\rm p}v_{\rm tang}}\,,
\end{equation}
where $v^\star$ is the target speed after the impact, $v$ is the projectile preimpact speed. The subscripts $norm$ and $tang$ denote the normal and tangential components to the target's local surface at the impact point, respectively. In our calculation, we use the experimentally obtained values of the efficiencies, but the thing is actually a bit more tricky. 

From the impact experiments we see that the efficiencies depend on impact angle $\sigma$, which is an angle between the impact velocity vector and the line perpendicular to the surface at the impact point. The experimental results give the following dependencies
\begin{equation}
  \eta(\sigma) = 1 + \eta_0\cos^2\sigma\qquad{\rm and}\qquad\zeta(\sigma)=\zeta_0\cos^2\sigma\,.
\end{equation}
\citet{yanagisawa2000} obtained the values of constants $\eta_0$ and $\zeta_0$ at $4\,\rm{km\,s^{-1}}$ impact speed, which is close to the median impact speed between the main belt bodies ($\sim 5\,\rm{km\,s^{-1}}$, \citealp{bottke2005}). For basaltic targets they give $\eta_0=1.52$ and $\zeta_0=0.409$. For sand target which could be similar to a regolith-covered asteroid, \citet{yanagisawa2002} gives $\zeta_0=0.687$. He did not obtain the $\eta_0$ value and therefore he uses $\eta_0=0$~($\eta=1$) for an asteroid large enough to reaccumulate all the ejecta by its gravity and we follow this approach.

There is also a relation between the efficiencies and the shape of the target body as described in \citet{yanagisawa2002}. It is sufficient to consider the equations given above since the impact angle $\sigma$ is different for a sphere and an ellipsoid for the same impact velocity and therefore we get different values of $\eta$ and $\zeta$ for different target shapes.

\subsection{Scaling laws}\label{scaling}

To find out how large will be the crater formed on the asteroid, we followed the scaling laws of \cite{holsapple-web} and \cite{holsapple1993}. The scaling laws in general predict the outcome of an experiment from another one with different values of problem parameters. When we study impact processes, we have centimetre-scale impact experiments from the lab and we try to predict the outcome of the same experiment on a much larger (kilometre) scale.

Based on shapes of simple craters on the Moon and other Solar System bodies we decided to model the impact crater as paraboloid of revolution. We assume that the aftermath of the crater formation (e.g., a collapse of the crater walls) does not have a substantial dynamical effect (the mass redistribution is small). 

The radius and depth of the crater is given by the following formulae
\begin{equation}
  R_{\rm c}=K_{\rm r}V_{\rm c}^{1/3}\qquad{\rm and}\qquad D_{\rm c}=K_{\rm d}V_{\rm c}^{1/3}\,,
\end{equation}
where $K_{\rm r}$ and $K_{\rm d}$ are the shape constants specific for a material of the asteroid (\citealp{holsapple-web}) and $V_{\rm c}$ is the volume of the crater given by
\begin{equation}
  V_{\rm c}=\frac{4\upi d^3}{3}\frac{\delta}{\rho}\pi_{\rm V}\,,
\end{equation}
$d$ being the impactor diameter, $\delta$ its density, $\rho$ is the target's density and $\pi_{\rm V}$ is given by
\begin{eqnarray}
\pi_{\rm V} = K_1\left\{\pi_2\left(\frac{\rho}{\delta}\right)^{(6\nu-2-\mu)/3\mu}\right. \nonumber \\ \nonumber
+ \left. K_2\left[\pi_3\left(\frac{\rho}{\delta}\right)^{(6\nu-2)/3\mu}\right]^{(2+\mu)/2}\right\}^{-3\mu/(2+\mu)} \\
\end{eqnarray}
\[
  \pi_2 = \frac{gd}{U^2}\,,\quad\pi_3=\frac{\bar{Y}}{\rho U^2}\,.\nonumber
\]
Constants $K_1$ and $K_2$ and exponents $\mu$ and $\nu$ come from laboratory impact experiments (\citealt{holsapple1993}), $g$ is the gravitational acceleration at the impact point, $U$ is the impact velocity component perpendicular to local surface and $\bar{Y}$ is the target's material strength. In our simulations we consider two cases, a rocky material corresponding to non-porous bodies and lunar regolith for porous asteroids. For the values we use, see Table~\ref{const_values}. 

\begin{table*}
\begin{minipage}{126mm}
 \caption{Scaling and material constants for non-porous and porous body materials.}
 \label{const_values}
 \begin{tabular}{c c c c c c c c c}
 \hline
 Material & $K_1$ & $K_2$ & $\mu$ & $\nu$ & $\bar{Y}\,(\rm{MPa})$ & \
 $\rho\,(\rm{g\,cm^{-3}})$ & $K_r$ & $K_d$ \\
 \hline
 Rock & 0.095 & 0.257 & 0.55 & 0.33 & 10 & 2.0 & 1.1 & 0.6\\
 Lunar regolith & 0.132 & 0.26 & 0.41 & 0.33 & 0.01 & 1.5 & 1.4 & 0.35\\
 \hline
 \end{tabular}
\end{minipage}
\end{table*}

\subsection{Synthetic lightcurves}\label{lightcurve}

We generated rotational lightcurves for the resulted model bodies. We used the program written by Mikko Kaasalainen (\citealt{kaasalainen2001}) and modified by Josef \v{D}urech (2011, pers. communication), which generates the lightcurve for an arbitrarily shaped body with given dynamical parameters. The shape is given by the set of oriented triangles building up the body surface mesh, the light scattering law was that of Hapke (\citealt{bowell1989}). The input also includes the orientation of axis of rotation at some instant of time as well as the Euler angles for that instant, that determine the body's orientation in space.

The Euler equations of motion are then solved and the integral light flux is calculated for given observational geometry with a chosen time increment. We simulated real observations with varying time sampling rates to avoid alias effects in generated lightcurves and we also added a Gaussian noise. 

\begin{table}
\begin{minipage}{126mm}
 \caption{Physical parameters of the sample target bodies.}
 \label{lightcurve_desc}
 \begin{tabular}{c c c c}
 \hline
 Figure & target mean & initial rotation & $\beta$ [deg] \\
        & diameter [km] & period [h] & \\
 \hline
 \ref{pa1} & 144.2 & 32 & 4.2 \\
 \ref{pa2} & 36.1 & 32 & 11.5 \\
 \ref{npa1} & 1.4 & 2 & 15.5 \\
 \ref{npa2} & 7.2 & 12 & 18.8 \\
 \ref{npa3} & 7.2 & 22 & 34.3 \\
 \hline
 \end{tabular}
\end{minipage}
\end{table}

\begin{figure}
        \centering
\includegraphics[angle=270,width=\columnwidth]{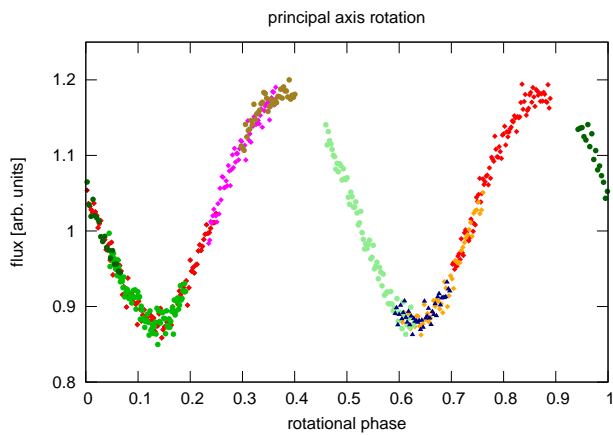}
\caption{A lightcurve of a body impacted by a small projectile, $\beta = 4.2^\circ$. The horizontal axis is the rotational phase, the vertical axis is a light flux normalised to unity. It is undistinguishable from a lightcurve of principal axis rotator.}
\label{pa1}
\end{figure}

\begin{figure}
        \centering
\includegraphics[angle=270,width=\columnwidth]{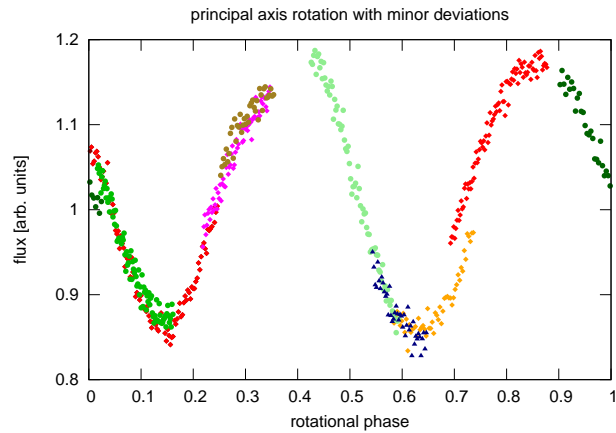}
\caption{A lightcurve with apparent minor deviations from single periodicity, $\beta = 11.5^\circ$, only marginally distinguishable by photometric observation.}
\label{pa2}
\end{figure}

\begin{figure}
        \centering
\includegraphics[angle=270,width=\columnwidth]{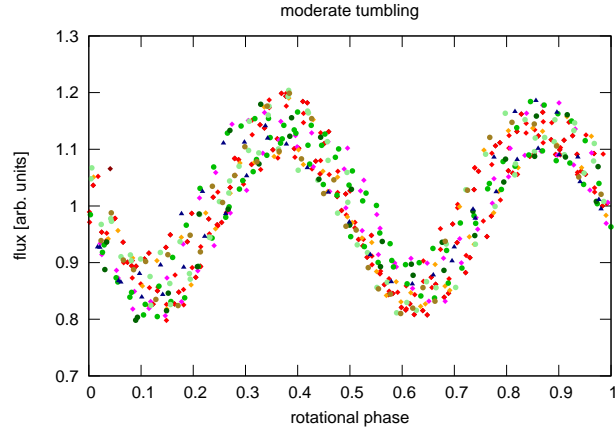}
\caption{A moderate tumbling signal is apparent in this lightcurve, $\beta = 15.5^\circ$.}
\label{npa1}
\end{figure}

\begin{figure}
        \centering
\includegraphics[angle=270,width=\columnwidth]{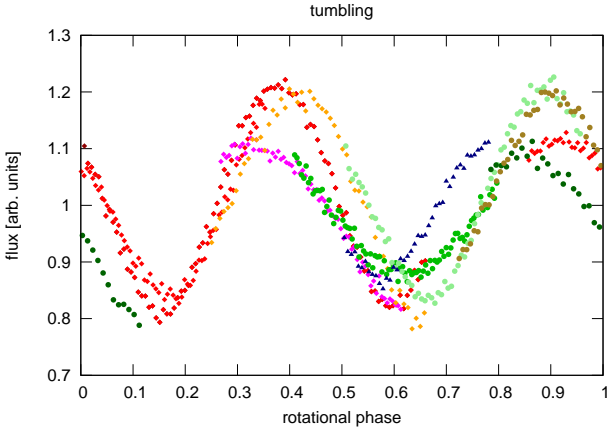}
\caption{In this lightcurve the tumbling is evident, $\beta = 18.8^\circ$.}
\label{npa2}
\end{figure}

\begin{figure}
        \centering
\includegraphics[angle=270,width=\columnwidth]{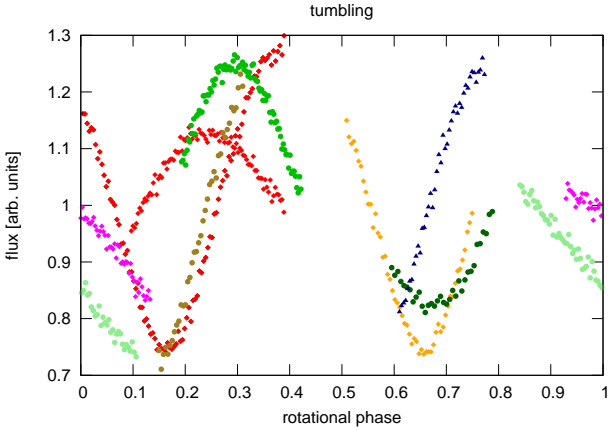}
\caption{Obvious tumbling for this slowly rotating body, $\beta = 34.3^\circ$.}
\label{npa3}
\end{figure}

A blind test lightcurve analysis was done for a set of cases with increasing $\beta$ to find out if the tumbling is detectable in the synthetic lightcurve. In Figs.~\ref{pa1}--\ref{npa3} we present five sample lightcurves. The axial ratios of the target were $a/c=2.0$ and $b/c=1.5$. The diameter of a projectile was about $5\%$ of the mean diameter of the target body in all these simulation runs. Other parameters describing the body are in Table~\ref{lightcurve_desc}.

The values of rotation axis misalignment for which the tumbling is clearly detected start at $\beta\sim 15^\circ$. In section~\ref{am_ratio} we will show that the value of $\beta$ is closely related to $L_{\rm{orb}}/L_{\rm t}$ ratio independent of other parameters describing the two bodies.

\section{Rotation axis misalignment}\label{rot_axis_mis}

In this section we present the results of our simulations. In individual runs, we varied one or two input parameters (e.g., body size or its rotation period) while keeping other constant. Then we plotted graphs showing the relationship between the varied parameters and the misalignment $\beta$ of the angular momentum vector after the collision (see section \ref{deviation}). The following graphs do not show results of all our simulations, but rather a representative sample.

The constant parameters describing the body in these runs were $a/c=2.0$, $b/c=1.5$, $\bar{Y}=10\,\rm{MPa}$, bulk densities of target and the projectile were $2\,\rm{g\,cm^{-3}}$. The projectile collided with the body at $v_x=-5\,\rm{km\,s^{-1}}$ and the impact point coordinates were $\phi=15^\circ$ and $\theta=15^\circ$. The incidence angle for such geometry is approximately $35^\circ$, which is ten degrees less than the statistical average for a spherical target body (\citealt{love1996}). 

\begin{figure}
        \centering
\includegraphics[angle=270,width=\columnwidth]{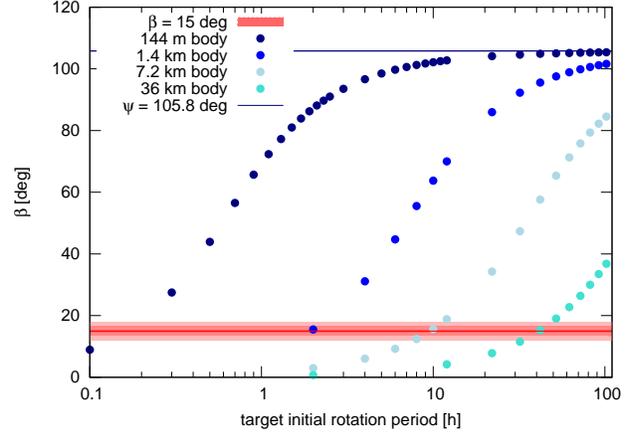}
\caption{A relationship between the AM vector misalignment and the initial rotation period of the body, for four different sizes. The projectile size was adjusted so that it excavated a crater of maximum size (see section~\ref{model}).}
\label{period}
\end{figure}

Figure~\ref{period} shows the misalignment of the AM vector for various values of the initial rotation period and four body sizes. As expected, $\beta$ increases with increasing rotation period. The slower rotation of the target implies it has lower angular momentum $L_{\rm t}$ and therefore the projectile with a given orbital angular momentum $L_{\rm{orb}}$ will excite the body more.

We can clearly see, that even for the largest body the tumbling could be detected for periods longer than about $42\,\rm{h}$. To understand qualitatively this graph, we can use dimensional analysis. For the rotational angular momentum of the target and for the orbital angular momentum of the projectile, respectively, we have 
\[
  L_{\rm t}\sim I_{\rm t}\omega_{\rm t}\sim m_{\rm t}r_{\rm t}^2\omega_{\rm t}\sim r_{\rm t}^5\frac{2\upi}{P_{\rm t}}\,,
\]
\begin{equation}
  L_{\rm{orb}}\sim r_{\rm t} m_{\rm p}v_{\rm{imp}}\,,
\end{equation}
where $r_{\rm t}$ is the mean target radius and $P_{\rm t}$ is the target's period of rotation. The ratio of these two is
\begin{equation}
  \frac{L_{\rm{orb}}}{L_{\rm t}}\sim\frac{P_{\rm t}}{r_{\rm t}^4}\,.
  \label{dim_anal_am}
\end{equation}
It means that the smaller the target and the longer the period of rotation, the higher the ratio of the angular momenta and therefore higher $\beta$. 

\begin{figure}
        \centering
\includegraphics[angle=270,width=\columnwidth]{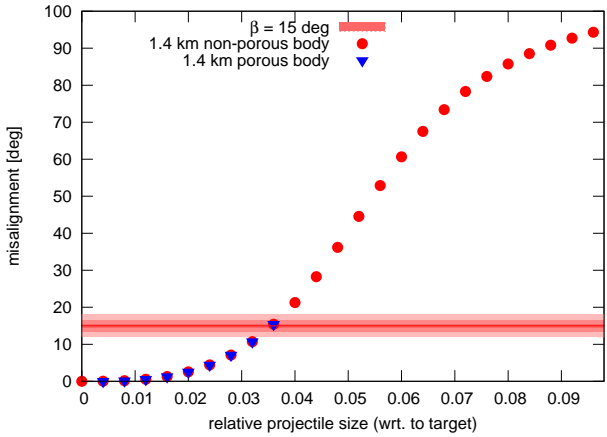}
\caption{A relationship between the AM vector misalignment and the projectile size for small target. The simulation for non-porous and porous body.}
\label{projectile_small}
\end{figure}

In another run, the sampled parameter was the projectile size. The results are plotted for small body both porous and non-porous (Fig.~\ref{projectile_small}). As the relative projectile size increases, $\beta$ also increases. A larger projectile makes a larger crater on the target body and hence larger change of the inertia tensor.

The parameters describing the body and the impact were the same as previously, the mean body diameter was $1.4\,\rm{km}$. For a porous target, there is a different set of the scaling and crater shape parameters as described in section~\ref{scaling} and different cratering efficiency (see section~\ref{efficiency}). There are only subtle differences of at most two degrees in $\beta$ between porous and non-porous targets.

\begin{figure}
        \centering
\includegraphics[angle=270,width=\columnwidth]{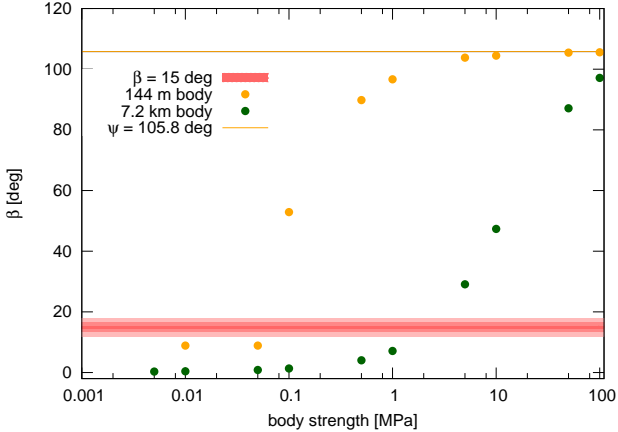}
\caption{A relationship between the AM vector misalignment and strength of the small and moderate size target.}
\label{strength}
\end{figure}

In Fig.~\ref{strength}, we plot the strength--misalignment relationship for a small and a moderate size body. For increasing target strength there is an increasing threshold crater size (see section~\ref{model} for details). We kept the actual crater size near the threshold, so for larger target strength we had larger projectile with larger $L_{\rm{orb}}$ ($L_{\rm t}$ being constant).

Other parameters than the strength and the body sizes ($144\,\rm{m}$ and $7.2\,\rm{km}$, respectively) were the same as in previous experiment, the initial rotation period was $32\,\rm{h}$.

\begin{figure}
        \centering
\includegraphics[angle=270,width=\columnwidth]{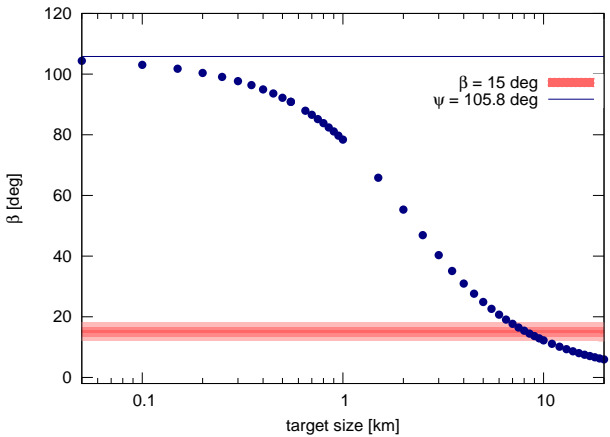}
\caption{A relationship between the AM vector misalignment and target size.}
\label{target_size}
\end{figure}

The last run shows a dependence of the misalignment on the target size, see Fig.~\ref{target_size}. It is clearly seen that with decreasing target size $\beta$ increases. As before, it was caused by increasing $L_{\rm{orb}}/L_{\rm t}$ ratio, which can be, again, explained by \nolinebreak di\nolinebreak{}men\nolinebreak{}sional \nolinebreak anal\nolinebreak{}ysis (see Eqn.~\ref{dim_anal_am}).

\section{Angular momenta ratio}\label{am_ratio}

Here we show, that the results of all our simulations can be described by a ratio of the projectile orbital to target rotational angular momentum $L_{\rm{orb}}/L_{\rm t}$. The relationship can be approximately described by the formula
\begin{equation}
  \cos\beta = \pm\left[1+\frac{\sin^2\psi}{(L_{\rm{t}}/L_{\rm orb}+\cos\psi)^2}\right]^{-1/2}\,,
\label{beta_am_ratio}
\end{equation}
where the $+$ sign is for $L_{\rm t}\geq -L_{\rm orb}\cos\psi$ and $-$ sign is for $L_{\rm t}<-L_{\rm orb}\cos\psi$, $\psi$ is the angle between the two angular mometum vectors just before the collision. For derivation of this formula see Appendix~\ref{app_beta}.

\begin{figure}
        \centering
\includegraphics[angle=270,width=\columnwidth]{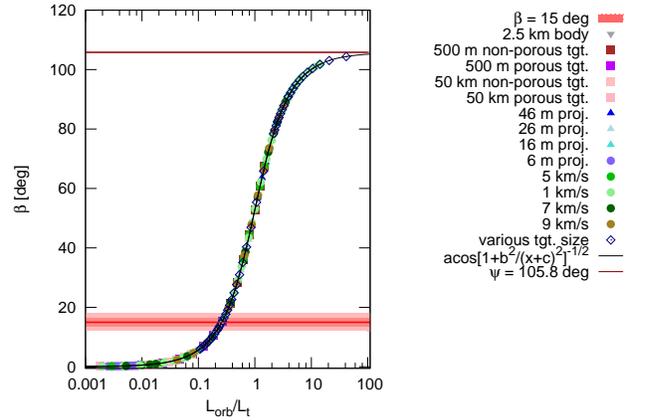}
\caption{All the previous results included in one graph, $\beta$ as a function of $L_{\rm{orb}}/L_{\rm t}$.}
\label{all}
\end{figure}

We plotted all the simulation results in Fig.~\ref{all} together with the curve given by Eqn.~\ref{beta_am_ratio}.

There is another obvious feature in the above mentioned graph\=the values of $\beta$ converge to the value of $\psi$. This is to be expected when the ratio of the magnitudes of these two vectors becomes much greater than one. 

\begin{figure}
        \centering
\includegraphics[angle=270,width=\columnwidth]{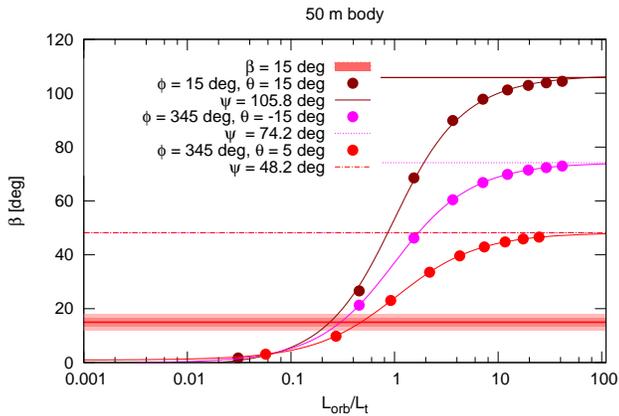}
\caption{Test of $\beta$ convergence to $\psi$ for various impact geometries, or angle between $\bmath{L_{\rm t}}$ and $\bmath{L_{\rm{orb}}}$ vectors.}
\label{beta_conv}
\end{figure}

We have checked this convergence also for other angles (other impact geometries), the results are presented in Fig.~\ref{beta_conv}. The target was $50\,\rm{m}$ body with other parameters as in previous simulations and the projectile collided with it at $v_x=-5\,\rm{km\,s^{-1}}$. The impact point coordinates are described in Fig. \ref{beta_conv} by $\phi$ and $\theta$ (see section~\ref{model}).

\section{Dispersal and shattering criteria}\label{disrupt_crit}

\begin{figure}
        \centering
\includegraphics[angle=270,width=\columnwidth]{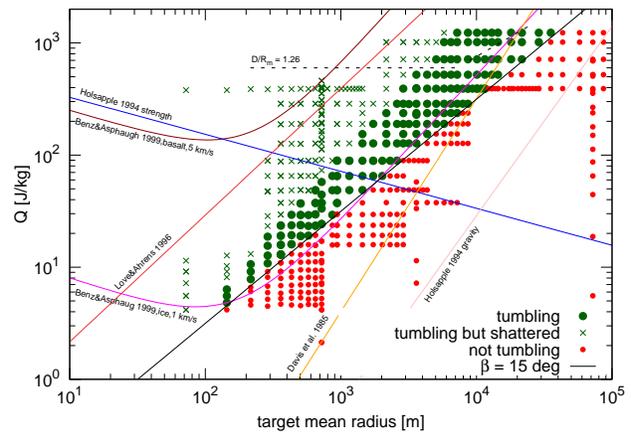}
\caption{The specific energy of collision as a function of the mean target radius. Small red circles denote bodies for which the tumbling was not apparent in the lightcurve. Large green circles are tumbling bodies which did not exceed the shattering criterion. Green cross symbols are for bodies that were shattered or even had a half of its mass dispersed by the collision. The dispersal criterion is according to \protect\cite{stewart2012}. Other dispersal criteria are also displayed. The thick black line is the limiting specific impact energy as derived in Appendix~\ref{border_specific}.}
\label{specific_energy_fig}
\end{figure}

It is important to compare the specific energy of the collision with the shattering or dispersal criterion so that the collision is subcatastrophic (cratering regime). Therefore we plotted specific impact energy as a function of the mean target radius and calculate the shattering criterion for most collisions we simulated. We considered targets with various dimensions, but constant $a/c=2.0$, $b/c=1.5$, with a rotation period of $36\,\rm{h}$, mean density $2\,\rm{g\,cm^{-3}}$ and strength $1\,\rm{MPa}$. The projectiles were of various sizes, their density was $2\,\rm{g\,cm^{-3}}$ and the impact velocity was $v_x=-5\,\rm{km\,s^{-1}}$. The impact point coordinates were $\phi=21^\circ$ and $\theta=21^\circ$, the incidence angle for such geometry is approximately $44^\circ$. The results are shown in Fig.~\ref{specific_energy_fig}. 

In the calculation, we followed \cite{stewart2012} who derived the dispersal criterion in gravity regime from their set of numerical simulations of collisions for wide variety of input parameters describing the colliding bodies. As shattering criterion we take $1/4$ of the dispersal value according to \cite{housen2009}. Since the diversity of dispersal criteria in literature is rather high, we plot other such functions in Fig.~\ref{specific_energy_fig} for comparison to our choice. 

We also plot the limiting value of the specific energy for which the tumbling after the collision becomes detectable in the lightcurve as described in section~\ref{lightcurve}. For its derivation see Appendix~\ref{border_specific}. The bodies, for which the specific impact energy of the collision was less than this limiting value, are labeled `not tumbling' for simplicity, but this label implies that the tumbling would not be detected in their lightcurves. If the collision energy exceeds the shattering criterion, we label the resulting body `tumbling but shattered' and we mark it with different symbol~($\times$) in the graph. 

We plotted an upper limit of the specific impact energy in the graph. It corresponds to the ratio of the diameter of the largest crater to the mean target radius $D/R_{\rm m}=1.26$ (the largest known crater on 253~Mathilde).

\section{Summary}\label{summary}

We studied effects of subcatastrophic collisions on asteroid rotation. We found that it is plausible that such collisions are responsible for observed excited rotation of the small bodies. The important parameter of the collision is the ratio of the projectile orbital angular momentum to the target rotational angular momentum before the collision ($L_{\rm{orb}}/L_{\rm t}$). We found an approximate relation between the rotational axis misalignment angle and this ratio, which can be used for estimation of the excitation effect of a particular collision between two bodies. 

We plotted the results of our simulations to the graph of the specific impact energy vs. mean target radius (Fig.~\ref{specific_energy_fig}). We showed that for a reasonable choice of input parameters and the dispersal limit, the excitation of rotation can take place below the shattering limit for a target body.

In a future work, we plan to estimate the probability of subcatastrophic collision capable of rotational excitation for bodies of various sizes during their lifetime. Also we will account for the damping timescale of tumbling asteroids to find out the probability of finding such bodies in non-principal axis rotation state. 

\section*{Acknowledgments}

We want to thank Josef \v{D}urech for kindly providing us with \verb+do_lcs_free+ program for generating the rotational lightcurves. We are also grateful to an anonymous reviewer for helpful comments. This work was supported by Masaryk University, Grant MUNI/A/0968/2009 and MUNI/A/0735/2012 and by Programme RVO~67985815.

\bibliographystyle{mn2e}
\bibliography{henych2013_mnras}

\appendix
\section{Missalignment angle\=angular momenta ratio dependence}\label{app_beta}

Equation~\ref{beta_main} defines the rotational axis misalignment angle after the collision. As we showed, it is closely related to the ratio of $L_{\rm{orb}}/L_{\rm t}$. Here we derive the approximate relation between these two quantities.

We assume that the shortest principal axis (represented by vector $\bmath{E_3}$) of the target remains unchanged during the collision (the small deviation is caused by the crater formed on the body). We only consider target rotational angular momentum $\bmath{L_{\rm t}}$ and projectile orbital angular momentum $\bmath{L_{\rm orb,p}}$ (further only $\bmath{L_{\rm orb}}$ for simplicity) and neglect all other angular momenta in Eqn.~\ref{am_conservation}, so the resulting $\bmath{L^\star}$ is

\begin{equation}
 \bmath{L^\star} = \bmath{L_{\rm orb}} + \bmath{L_{\rm t}}\,.
\end{equation}

Hence we have
\begin{equation}
  \cos\beta = \frac{\bmath{L^{\!\star}}\bmath{E_3}}{|\bmath{L^{\!\star}}||\bmath{E_3}|}=\frac{(\bmath{L_{\rm orb}} + \bmath{L_{\rm t}})_z}{|\bmath{L_{\rm orb}} + \bmath{L_{\rm t}}|}\,,
\label{cos_beta_app_def}
\end{equation}
where the subscript $z$ denotes the $z$ component of the vector. This can be further written as
\begin{equation}
  \cos\beta = \frac{(L_{\rm orb}\cos\psi + L_{\rm t})}{[(L_{\rm orb}\cos\psi + L_{\rm t})^2 + L^2_{\rm orb}\sin^2\psi]^{1/2}}\,,
\label{cos_beta_app_exp}
\end{equation}
where $\psi$ is the angle between $\bmath{L_{\rm orb}}$ and $\bmath{L_{\rm t}}$ vectors. After some algebra we have
\begin{equation}
  \cos\beta = \pm\left[1+\frac{\sin^2\psi}{(L_{\rm{t}}/L_{\rm orb}+\cos\psi)^2}\right]^{-1/2}\,,
\label{cos_beta_app}
\end{equation}
where the $+$ sign is for $L_{\rm t}\geq -L_{\rm orb}\cos\psi$ and $-$ sign is for $L_{\rm t}<-L_{\rm orb}\cos\psi$.

\section{Limiting specific impact energy}\label{border_specific}

Here we derive the limiting value of the specific impact energy $Q^{\rm tumb}$ as a function of physical parameters of the colliding bodies and the ratio of their angular momenta (orbital angular momentum of the projectile $L_{\rm orb}$ and the rotational angular momentum of the target before the collision $L_{\rm t}$).

The ratio of the angular momenta is given by 
\begin{equation}
  \frac{L_{\rm t}}{L_{\rm orb}}=\frac{2\upi (a^2+b^2)abc\rho}{5r^3_{\rm p}v_{\rm imp}\delta P_{\rm t}r_{\rm t}}\,,
\label{am_ratio_eqn}
\end{equation}
where $v_{\rm imp}$ is the impact speed, $\delta$ and $\rho$ are the projectile and target density, respectively, $r_{\rm t}$ and $r_{\rm p}$ are their mean radii, respectively, $a$, $b$ and $c$ are the target's semi-axes and $P_{\rm t}$ is the target's period of rotation. The specific energy of the collision in the system's centre of mass for $m_{\rm p}\ll m_{\rm t}$ is
\begin{equation}
  Q=\frac{m_{\rm p}v^2_{\rm imp}}{2m_{\rm t}}=\frac{r^3_{\rm p}v^2_{\rm imp}\delta}{abc\rho}\,,
\label{specific_energy_eqn}
\end{equation}
$ m_{\rm t}$ and $m_{\rm p}$ being the target and the projectile mass, respectively. 

We solve Eqn.~\ref{cos_beta_app} for $L_{\rm t}/L_{\rm orb}$ ratio, Eqn.~\ref{am_ratio_eqn} for $r_{\rm p}$ and insert both of them into Eqn.~\ref{specific_energy_eqn}. We then obtain the specific energy for which the tumbling starts to be apparent in the lightcurve
\begin{equation}
  Q^{\rm tumb}=\frac{\upi(a^2+b^2)v_{\rm imp}}{5 r_{\rm t}P_{\rm t}}\left(\frac{\sin^2\psi}{\cos^{-2}\beta-1}-\cos^2\psi\right)^{-1}\,.
\label{specific_energy}
\end{equation}
In Fig.~\ref{specific_energy_fig} we use the derived value of $\beta=15^\circ$.

\bsp

\label{lastpage}

\end{document}